# Bianchi type-II universe with wet dark fluid in General Theory of Relativity

**Chandra Rekha Mahanta** and **Azizur Rahman Sheikh**

Department of Mathematics, Gauhati University, Guwahati-781014 (INDIA)

Email. crmahanta4@gmail.com, azizur08@gmail.com

**Abstract:** In this paper, dark energy models of the universe filled with wet dark fluid are constructed in the frame work of LRS Bianchi type-II space-time in General Theory of Relativity. A new equation of state modeled on the equation of state $p = \gamma(\rho - \rho_*)$, which can describe liquid including water, is used. The exact solutions of Einstein's field equations are obtained in quadrature form and the models corresponding to the cases γ = 0 and γ = 1 are discussed in details.

## 1   Introduction

The studies of remote type Ia supernova (SNeIa) in 1998 ( Perlmutter et al. 1998,1999, Riess et al. 1998, 2004 ) suggest that the expansion of the universe is accelerating. This discovery is viewed as a major breakthrough of the observational cosmology as till 1998 it was thought that the universe was expanding with deceleration due to the attraction of the masses within it. The accelerated expansion of the universe can be accounted for by attributing 68.3% (Ade et al. 2013) of the energy density of the universe to the a mysterious source of energy dubbed as dark energy ( DE ) with negative pressure, the true nature of which is still unknown. However, there is no dearth of candidates for DE proposed in literature. Some of them are the cosmological constant (Λ), Quintessence (Ratra et al. 1988; Caldwell et al.1998 ), K-essence ( Armendariz-Picon et al. 1999, 2001 ), phantom energy ( Caldwell 2002, 2003; Carrol et al. 2003 ), Chaplygin gas (Kamenshchik et al. 2001 ) etc.. Holman and Naidu (2005) introduced a new candidate for DE, called the wet dark fluid (WDF) with the equation of state

$$p_w = \gamma(\rho_w - \rho_*) \tag{1}$$

where $p_w$ is the pressure, $\rho_w$ the energy density of WDF and the parameters $\gamma$ and $\rho_*$ are taken to be positive with the restriction $0 < \gamma < 1$. Babichev et al. (2005) also proposed a dark energy model with a linear equation of state similar to (1), which is $p = \alpha(\rho - \rho_0)$, where $\alpha$ and $\rho_0$ are free parameters, to overcome the hydrodynamical instability of the dark energy with the usually used EOS $p = \omega\rho$ where $\omega =$ constant $< 0$. Eqn. (1) is modeled on an empirical EOS proposed by Tait (1988) and Hayword (1967) to treat water and aqueous solutions. WDF has two components: one of them behaves as the Cosmological constant and the other as standard fluid including water. The second component can be used as dark matter. Thus, WDF unifies the two dark components. Many authors like Singh and Chaubey (2008), Adhav (2010), Chaubey (2009,2011 ), Katore et al. ( 2012), Ravishankar (2013 ), Chirde and Rahate ( 2013 ), Deo et al.(2014 ), Kandalkaret al. ( 2014 ) studied WDF in different contexts.

Though it is believed that our universe is spatially flat, some experimental data have suggested that our present universe is not perfectly flat, but it possesses a small curvature (Bennet et al. 2003; Spergel et al. 2003).In this scenario, Bianchi type metrics are suitable for studying the evolution of the universe as these give anisotropy and ellipsoidality to the universe. However, the Bianchi models isotropize at late times even for ordinary matter. In fact this isotropization is due to the implicit assumption that the DE is isotropic in nature. Bianchi type - II space-time models can be used to construct cosmological models suitable for describing early stages of evolution of the universe. Recently, Pradhan et al. (2011), Kumar and Akarsu (2011), Singh and Kumar (2006), Saha and Yadav (2012 ) have studied Bianchi type-II models in different contexts.

In this paper, we present cosmological models in the framework of LRS Bianchitype-II space-time with wet dark fluid. We have used the EOS (1) for WDF with the restriction $0 \leq \gamma \leq 1$. The paper is organized as follows: In Sec.2 we obtain field equations for LRS Bianchi type-II metric and in Sec.3, we obtain exact solutions for the field equations in quadrature form. In Sec.4, we discuss some particular cases of the obtained results together with their physical properties. We conclude the paper with a brief discussion in Sec. 5.

## 2 Metric and field equations

We consider anisotropic LRS Bianchi type-II metric in the form

$$ds^2 = -dt^2 + A^2 ( dx - zdy )^2 + B^2 ( dy^2 + dz^2 ) \qquad (2)$$

where *A* and *B* are directional scale factors and are functions of cosmic time *t* alone.

The Einstein's field equations are given by

$$R_{ij} - \frac{1}{2} R g_{ij} = k T_{ij} \qquad (3)$$

where $k$ is the gravitational constant and $T_{ij}$ is the energy-momentum tensor for WDF and is given by

$$T_i^j = (\rho_w + p_w) u_i u^j - p_w g_i^j \qquad (4)$$

where $u^i$ is the flow vector satisfying

$$g_{ij}u^i u^j = 1 \quad (5)$$

In comoving co-ordinates, the field equations (3) with respect to the metric (2) are found to be

$$2\frac{\dot{A}\dot{B}}{AB} + \left(\frac{\dot{B}}{B}\right)^2 - \frac{1}{4}\frac{A^2}{B^4} = k\rho_w \tag{6}$$

$$\frac{\ddot{A}}{A} + \frac{\ddot{B}}{B} + \frac{\dot{A}\dot{B}}{AB} + \frac{1}{4}\frac{A^2}{B^4} = -kp_w \tag{7}$$

$$2\frac{\ddot{B}}{B} + \left(\frac{\dot{B}}{B}\right)^2 - \frac{3}{4}\frac{A^2}{B^4} = -kp_w \tag{8}$$

where an over dot denotes differentiation with respect to $t$.

The equation of continuity is

$$\dot{\rho}_w + 3H(\rho_w + p_w) = 0 \tag{9}$$

where $H$ is the average Hubble parameter given by

$$3H = \frac{\dot{V}}{V} = \frac{\dot{A}}{A} + 2\frac{\dot{B}}{B} \tag{10}$$

The volume expansion parameter $V$ is given by

$$V = AB^2 \tag{11}$$

The expansion scalar $\theta$, anisotropy parameter $\Delta$, shear scalar $\sigma^2$ and the deceleration parameter $q$, which is a dimensionless measure of the cosmic acceleration of the expansion of the universe, are defined by

$$\theta = \frac{\dot{V}}{V} = \frac{\dot{A}}{A} + 2\frac{\dot{B}}{B} \tag{12}$$

$$\Delta = \frac{1}{3}\sum_{i=1}^{3}\frac{[H_i - H]^2}{H^2} = \frac{2}{3H^2}\left(\frac{\dot{A}}{A} - \frac{\dot{B}}{B}\right)^2 \tag{13}$$

$$\sigma^2 = \frac{1}{2}[\sum_{i=1}^{3} H_i^2 - \frac{1}{3}\theta^2] = \frac{3}{2}\Delta H^2 \tag{14}$$

$$q = \frac{d}{dt}\left(\frac{1}{H}\right) - 1 \tag{15}$$

Again, using Eqn. (1) in (9) with $H = \frac{1}{3}\frac{\dot{V}}{V}$ and integrating we get

$$\rho_w = \frac{C}{V^{1+\gamma}} + \frac{\gamma}{1+\gamma}\rho_* \tag{16}$$

where $C(> 0)$ is a constant of integration.

## 3 The solution of the field equations:

From Eqns. (6)—(8), we have

$$\frac{\ddot{A}}{A} + 2\frac{\ddot{B}}{B} + 4\frac{\dot{A}\dot{B}}{AB} + 2\left(\frac{\dot{B}}{B}\right)^2 - \frac{A^2}{2B^4} = \frac{3k}{2}(\rho_w - p_w) \tag{17}$$

Again, from Eqn. (10)

$$\frac{\ddot{V}}{V} = \frac{\ddot{A}}{A} + 2\frac{\ddot{B}}{B} + 4\frac{\dot{A}\dot{B}}{AB} + 2\left(\frac{\dot{B}}{B}\right)^2 \tag{18}$$

Using Eqn. (18) in Eqn. (17), we get

$$\frac{\ddot{V}}{V} - \frac{A^2}{2B^4} = \frac{3k}{2}(\rho_w - p_w) \tag{19}$$

Eqns. (9) and (19) with $H = \frac{1}{3}\frac{\dot{V}}{V}$ yield

$$\frac{\dot{V}\ddot{V}}{V^2} - \frac{A^2}{2B^4}\frac{\dot{V}}{V} - \frac{3k}{2}\dot{\rho}_w = 3k\frac{\dot{V}}{V}\rho_w \tag{20}$$

which can be put in the form

$$2\dot{V}\ddot{V} - \frac{A^2}{B^4}V\dot{V} = 3k\frac{d}{dt}(V^2 \rho_w) \tag{21}$$

Now, we have three equations (6) - (8) and four unknown parameters $A$, $B$, $\rho_w$ and $p_w$. So, we require one more physical condition relating the parameters to solve the equations completely. For this purpose, we take

$$B = A^m \quad (m > 0) \tag{22}$$

From Eqns. (11) and (22) we get

$$A = V^{\frac{1}{2m+1}}, \quad B = V^{\frac{m}{2m+1}} \tag{23}$$

Eqns. (21) and (23) yield

$$2\dot{V}\ddot{V} - V^{\frac{3-2m}{2m+1}}\dot{V} = 3k\frac{d}{dt}(V^2 \rho_w) \tag{24}$$

Again, from Eqns. (7) and (8)

$$\frac{\ddot{A}}{A} - \frac{\ddot{B}}{B} + \frac{\dot{A}\dot{B}}{AB} - \left(\frac{\dot{B}}{B}\right)^2 + \frac{A^2}{B^4} = 0 \tag{25}$$

Using Eqn. (23) in Eqn. (25) we get

$$\ddot{V} + \frac{2m+1}{1-m} V^{\frac{3-2m}{2m+1}} = 0, \qquad (m \neq 1) \tag{26}$$

which yields

$$V^{\frac{3-2m}{2m+1}} = \frac{m-1}{2m+1} \ddot{V} \tag{27}$$

Using Eqn. (27) in Eqn.(24) we get

$$2\dot{V}\ddot{V} = \frac{2k(2m+1)}{m+1} \frac{d}{dt}(V^2 \rho_w) \tag{28}$$

On integration, Eqn. (28) yields

$$\dot{V}^2 = C_2 \rho_w V^2 + C_1 \tag{29}$$

where $C_1$ is the integration constant and

$$C_2 = \frac{2k(2m+1)}{m+1} \tag{30}$$

Eqn. (29) with Eqn.(16) yields

$$\int \frac{dV}{\sqrt{C_2 \left( \frac{C}{V^{1+\gamma}} + \frac{\gamma}{1+\gamma} \rho_* \right) V^2 + C_1}} = t + t_0 \tag{31}$$

where $t_0$ is the integration constant. We may take $t_0 = 0$ as this will only shift the origin of time.

### 4   Some particular cases:

*Case I*: $\gamma = 0$ ( Dust universe )

Eqn. (31) reduces to

$$\int \frac{dV}{\sqrt{CC_2 V + C_1}} = t \tag{32}$$

which yields

$$V = \frac{CC_2}{4} t^2 - \frac{C_1}{CC_2} \tag{33}$$

$$A = \left( \frac{CC_2}{4} t^2 - \frac{C_1}{CC_2} \right)^{\frac{1}{2m+1}} \tag{34}$$

$$B = \left( \frac{CC_2}{4} t^2 - \frac{C_1}{CC_2} \right)^{\frac{m}{2m+1}} \tag{35}$$

$$\frac{\dot{V}}{V} = \frac{2t}{t^2 - \frac{4C_1}{c^2 c_2^2}} \tag{36}$$

$$\frac{\dot{A}}{A} = \frac{2t}{(2m+1)\left(t^2 - \frac{4C_1}{c^2 c_2^2}\right)} \tag{37}$$

$$\frac{\dot{B}}{B} = \frac{2mt}{(2m+1)\left(t^2 - \frac{4C_1}{c^2 c_2^2}\right)} \tag{38}$$

$$H = \frac{2t}{3\left(t^2 - \frac{4C_1}{c^2 c_2^2}\right)} \tag{39}$$

$$\theta = \frac{2t}{t^2 - \frac{4C_1}{c^2 c_2^2}} \tag{40}$$

$$\Delta = \frac{6(m-1)^2}{(2m+1)^2} \tag{41}$$

$$\sigma^2 = \frac{4(m-1)^2 t^2}{(2m+1)^2 \left(t^2 - \frac{4C_1}{c^2 c_2^2}\right)^2} \tag{42}$$

$$\rho_w = \frac{4C}{CC_2 t^2 - \frac{4C_1}{CC_2}} \tag{43}$$

$$p_w = 0 \tag{44}$$

$$q = \frac{1}{2} + \frac{6C_1}{c^2 c_2^2 t^2} \tag{45}$$

Since, $V$ is positive at $t = 0$, therefore, $C_1 < 0$.

We take

$$C_1 = -C_3 , C_3 > 0 \tag{46}$$

$A$, $B$, $V$ are all finite at $t = 0$ and infinite at $t = \infty$.

$\rho_w$ is finite at $t = 0$ and decreases to zero as $t$ increases to $\infty$.

Since, $\Delta \neq 0$ ( for $m \neq 1$), the universe is anisotropic throughout its evolution. Also since q < 0 when $t < \frac{2\sqrt{3C_3}}{CC_2}$ and q > 0 when $> \frac{2\sqrt{3C_3}}{CC_2}$, the universe expands with acceleration when $t < \frac{2\sqrt{3C_3}}{CC_2}$ and with deceleration when $t > \frac{2\sqrt{3C_3}}{CC_2}$.

*Case II*: $\gamma = 1$ ( Zeldovich fluid )

Eqn. (31) reduces to

$$\int \frac{dV}{\sqrt{C_1 + CC_2 + \frac{C_2\rho_*}{2}V^2}} = t \tag{47}$$

which yields

$$V = e^{\sqrt{\frac{C_2\rho_*}{2}}t}, \text{ if } C_1 + CC_2 = 0 \tag{48}$$

$$V = \sqrt{\frac{2(C_1 + CC_2)}{C_2\rho_*}}\sinh\left(\sqrt{\frac{C_2\rho_*}{2}}t\right), \text{ if } C_1 + CC_2 > 0 \tag{49}$$

$$V = \sqrt{\frac{-2(C_1 + CC_2)}{C_2\rho_*}}\cosh\left(\sqrt{\frac{C_2\rho_*}{2}}t\right), \text{ if } C_1 + CC_2 < 0 \tag{50}$$

***Subcase II (a)***: When $C_1 + CC_2 = 0$, we have

$$V = e^{\sqrt{\frac{C_2\rho_*}{2}}t} \tag{51}$$

$$A = e^{\frac{\sqrt{\frac{C_2\rho_*}{2}}t}{2m+1}} \tag{52}$$

$$B = e^{\frac{m\left(\sqrt{\frac{C_2\rho_*}{2}}t\right)}{2m+1}} \tag{53}$$

$$\frac{\dot{V}}{V} = \sqrt{\frac{C_2\rho_*}{2}} \tag{54}$$

$$\frac{\dot{A}}{A} = \frac{\sqrt{\frac{C_2\rho_*}{2}}}{2m+1} \tag{55}$$

$$\frac{\dot{B}}{B} = \frac{m\sqrt{\frac{C_2\rho_*}{2}}}{2m+1} \tag{56}$$

$$H = \frac{1}{3}\sqrt{\frac{C_2\rho_*}{2}} \tag{57}$$

$$\theta = \sqrt{\frac{C_2\rho_*}{2}} \tag{58}$$

$$\Delta = \frac{6(m-1)^2}{(2m+1)^2} \tag{59}$$

$$\sigma^2 = \frac{C_2\rho_*(m-1)^2}{2(2m+1)^2} \tag{60}$$

$$\rho_w = \frac{C}{e^{\sqrt{2C_2\rho_* t}}} + \frac{\rho_*}{2} \tag{61}$$

$$p_w = \frac{C}{e^{\sqrt{2C_2\rho_* t}}} - \frac{\rho_*}{2} \tag{62}$$

$$q = -1 \tag{63}$$

$A, B, V$ are all unity at $t = 0$ and infinite at $t = \infty$.

$\rho_w$ is finite at $t = 0$ and decreases to $\frac{\rho_*}{2}$ as $t$ increases to $\infty$. $p_w$ is finite at $t = 0$ and decreases to $-\frac{\rho_*}{2}$ as $t$ increases to $\infty$.

Since, $\Delta \neq 0$ ( for $m \neq 1$), the universe is anisotropic throughout its evolution.

Again, since, $q < 0$, the universe expands with acceleration throughout its evolution.

Since, $\frac{dH}{dt} = 0$, in this case the universe has the fastest rate of expansion.

***Subcase II (b)***: For $C_1 + CC_2 > 0$, we have

$$V = \sqrt{\frac{2(C_1 + CC_2)}{C_2\rho_*}} \sinh\left(\sqrt{\frac{C_2\rho_*}{2}}\, t\right) \tag{64}$$

$$A = \left[\sqrt{\frac{2(C_1 + CC_2)}{C_2\rho_*}} \sinh\left(\sqrt{\frac{C_2\rho_*}{2}}\, t\right)\right]^{\frac{1}{2m+1}} \tag{65}$$

$$B = \left[\sqrt{\frac{2(C_1 + CC_2)}{C_2\rho_*}} \sinh\left(\sqrt{\frac{C_2\rho_*}{2}}\, t\right)\right]^{\frac{m}{2m+1}} \tag{66}$$

$$\frac{\dot{V}}{V} = \sqrt{\frac{C_2\rho_*}{2}} \coth\left(\sqrt{\frac{C_2\rho_*}{2}}\, t\right) \tag{67}$$

$$\frac{\dot{A}}{A} = \frac{\sqrt{\frac{C_2\rho_*}{2}}}{2m+1} \coth\left(\sqrt{\frac{C_2\rho_*}{2}}\, t\right) \tag{68}$$

$$\frac{\dot{B}}{B} = \frac{m\sqrt{\frac{C_2\rho_*}{2}}}{2m+1} \coth\left(\sqrt{\frac{C_2\rho_*}{2}}\, t\right) \tag{69}$$

$$H = \frac{1}{3}\sqrt{\frac{C_2\rho_*}{2}} \coth\left(\sqrt{\frac{C_2\rho_*}{2}}\, t\right) \tag{70}$$

$$\theta = \sqrt{\frac{C_2\rho_*}{2}}\coth\left(\sqrt{\frac{C_2\rho_*}{2}}\,t\right) \tag{71}$$

$$\Delta = \frac{6(m-1)^2}{(2m+1)^2} \tag{72}$$

$$\sigma^2 = \frac{C_2\rho_*(m-1)^2}{2(2m+1)^2}\coth^2\left(\sqrt{\frac{C_2\rho_*}{2}}\,t\right) \tag{73}$$

$$\rho_w = \frac{CC_2\rho_*}{2(C_1+CC_2)}\operatorname{cosech}^2\left(\sqrt{\frac{C_2\rho_*}{2}}\,t\right) + \frac{\rho_*}{2} \tag{74}$$

$$p_w = \frac{CC_2\rho_*}{2(C_1+CC_2)}\operatorname{cosech}^2\left(\sqrt{\frac{C_2\rho_*}{2}}\,t\right) - \frac{\rho_*}{2} \tag{75}$$

$$q = 3\operatorname{sech}^2\left(\sqrt{\frac{C_2\rho_*}{2}}\,t\right) - 1 \tag{76}$$

$A, B, V$ are all zero at $t = 0$ and infinite at $t = \infty$.

$\rho_w$ is infinite at $t = 0$ and decreases to $\frac{\rho_*}{2}$ as $t$ increases to $\infty$. $p_w$ is infinite at $t = 0$ and decreases to $-\frac{\rho_*}{2}$ as $t$ increases to $\infty$.

Since, $\Delta \neq 0$ ( for $m \neq 1$), the universe is anisotropic throughout its evolution.

Again, since, $q < 0$ when $t > \sqrt{\frac{2}{C_2\rho_*}}\ln(\sqrt{3}+\sqrt{2})$, the universe expands with acceleration at late times.

***Subcase II (c)***: For $C_1 + CC_2 < 0$, we have

$$V = \sqrt{\frac{-2(C_1+CC_2)}{C_2\rho_*}}\cosh\left(\sqrt{\frac{C_2\rho_*}{2}}\,t\right) \tag{77}$$

$$A = \left[\sqrt{\frac{-2(C_1+CC_2)}{C_2\rho_*}}\cosh\left(\sqrt{\frac{C_2\rho_*}{2}}\,t\right)\right]^{\frac{1}{2m+1}} \tag{78}$$

$$B = \left[\sqrt{\frac{-2(C_1+CC_2)}{C_2\rho_*}}\cosh\left(\sqrt{\frac{C_2\rho_*}{2}}\,t\right)\right]^{\frac{m}{2m+1}} \tag{79}$$

$$\frac{\dot{V}}{V} = \sqrt{\frac{C_2\rho_*}{2}}\tanh\left(\sqrt{\frac{C_2\rho_*}{2}}\,t\right) \tag{80}$$

$$\frac{\dot{A}}{A} = \frac{\sqrt{\frac{C_2 \rho_*}{2}}}{2m+1} \tanh\left(\sqrt{\frac{C_2 \rho_*}{2}} t\right) \tag{81}$$

$$\frac{\dot{B}}{B} = \frac{m}{2m+1} \sqrt{\frac{C_2 \rho_*}{2}} \tanh\left(\sqrt{\frac{C_2 \rho_*}{2}} t\right) \tag{82}$$

$$H = \frac{1}{3} \sqrt{\frac{C_2 \rho_*}{2}} \tanh\left(\sqrt{\frac{C_2 \rho_*}{2}} t\right) \tag{83}$$

$$\theta = \sqrt{\frac{C_2 \rho_*}{2}} \tanh\left(\sqrt{\frac{C_2 \rho_*}{2}} t\right) \tag{84}$$

$$\Delta = \frac{6(m-1)^2}{(2m+1)^2} \tag{85}$$

$$\sigma^2 = \frac{C_2 \rho_* (m-1)^2}{2(2m+1)^2} \tanh^2\left(\sqrt{\frac{C_2 \rho_*}{2}} t\right) \tag{86}$$

$$\rho_w = \frac{CC_2 \rho_*}{2(C_1 + CC_2)} \text{sech}^2\left(\sqrt{\frac{C_2 \rho_*}{2}} t\right) + \frac{\rho_*}{2} \tag{87}$$

$$p_w = \frac{CC_2 \rho_*}{2(C_1 + CC_2)} \text{sech}^2\left(\sqrt{\frac{C_2 \rho_*}{2}} t\right) - \frac{\rho_*}{2} \tag{88}$$

$$q = -3\text{cosech}^2\left(\sqrt{\frac{C_2 \rho_*}{2}} t\right) - 1 \tag{89}$$

$A, B, V$ are all finite at $t = 0$ and infinite at $t = \infty$.

$\rho_w$ is finite at $t = 0$ and increases to $\frac{\rho_*}{2}$ as $t$ increases to $\infty$. $p_w$ is finite at $t = 0$ and increases to $-\frac{\rho_*}{2}$ as $t$ increases to $\infty$. Also, since, $\rho_w > 0$ at $t = 0$, hence, $C_1 + 2CC_2 < 0$.

Since, $\Delta \neq 0$ ( for $m \neq 1$), the universe is anisotropic throughout its evolution.

Again, since, $q < 0$ for all $t$, the universe expands with acceleration throughout its evolution.

## 5 Conclusion

We have studied the universe filled with an isotropic wet dark fluid (WDF) obeying the equation of state $p_w = \gamma(\rho_w - \rho_*)$ with $0 \leq \gamma \leq 1$ in the frame-work of LRS Bianchi type-II space-time in General Relativity. The solution of the field equations is obtained in quadrature form. We consider four particular cases corresponding to $\gamma = 0$ and $\gamma = 1$ in detail and discuss some physical properties of the universe represented by the models. It is found that in all the models the universe is anisotropic throughout its evolution though the WDF is isotropic. We have also found that:

(i) For $\gamma = 0$, the universe contains non-phantom energy throughout its evolution. Again, since $q < 0$ when $t < \frac{2\sqrt{3C_3}}{CC_2}$ and $q > 0$ when $t > \frac{2\sqrt{3C_3}}{CC_2}$, the universe expands with acceleration for some time after the big bang and then expands with retardation till it reaches the big crunch.

This model is suitable for representing the inflationary universe of the early era and the universe of the matter dominating era.

(ii) For $\gamma = 1$ and $C_1 + CC_2 = 0$, since $\omega_w \geq -1$, the universe contains non-phantom energy. Also, $\omega_w = -1$ at $= \infty$, hence in this case, the WDF behaves like cosmological constant at late times of the universe. Again, since $q < 0$ for all $t$, the universe, after its start from the big-bang, expands with acceleration throughout its evolution.

This model is suitable for early and late time accelerated expanding universe.

(iii) For $\gamma = 1$ and $C_1 + CC_2 > 0$, since $\omega_w \geq -1$, the universe contains non-phantom energy. Also, $\omega_w = -1$ at $t = \infty$, hence in this case, the WDF behaves like cosmological constant at late times of the universe. Again, since $q < 0$ for large $t$, this model is suitable for representing the late time accelerated expanding universe.

(iv) For $\gamma = 1$ and $C_1 + CC_2 < 0$, since $\omega_w \leq -1$, the universe contains phantom energy. Also, $\omega_w = -1$ at $t = \infty$, hence the WDF behaves like cosmological constant at late times of the universe. Again, since $q < 0$ for all $t$, this model is suitable for representing the early time as well as the late time accelerated expanding universe.

Acknowledgements:

The authors thank the unknown referee for very valuable comments and suggestions.